\documentclass[amssymb,useAMS,prd,aps,twocolumn,superscriptaddress]{revtex4}
\usepackage{pslatex}
\usepackage{graphicx}
\usepackage{psfrag}

\def\refe@jnl#1{{#1}}
\def\aj{\refe@jnl{Astron.~J.}}
\def\araa{\refe@jnl{Annu.~Rev.~Astron.~Astrophys.}}
\def\apj{\refe@jnl{Astrophys.~J.}}
\def\apjl{\refe@jnl{Astrophys.~J.~Lett.}}
\def\aap{\refe@jnl{Astron.~Astrophys.}}
\def\mnras{\refe@jnl{Mon.~Not.~R.~Astron.~Soc.}}
\def\prd{\refe@jnl{Phys.~Rev.~D}}
\def\fcp{\refe@jnl{Fund.~Cos.~Phys.}}
\def\physrep{\refe@jnl{Phys.~Rep.}}
\def\physlett{\refe@jnl{Phys.~Lett.}}

\def\invisible#1{  }

\def\lsim{\mathrel{\lower4pt\hbox{$\sim$}} 
\hskip-9.5pt\raise1.6pt\hbox{$<$}\;} 
 
\def\gsim{\mathrel{\lower4pt\hbox{$\sim$}} 
\hskip-9.5pt\raise1.6pt\hbox{$>$}\;}

\begin{document}

\title{Electroweak Symmetry Breaking induced by  Dark Matter}

\author{Thomas Hambye and Michel H.G. Tytgat}
\affiliation{Service de Physique Th\'eorique\\
 Universit\'e Libre de Bruxelles CP225\\
Boulevard du Triomphe, 1050 Brussels, Belgium}


\begin{abstract}

The mechanism behind Electroweak Symmetry Breaking (EWSB) and the nature of dark matter (DM) are currently among the most important 
issues in high energy  physics. Since a natural 
dark matter candidate is a weakly interacting massive particle or WIMP, 
with mass around the electroweak scale, 
it is clearly of interest to investigate the possibility that DM and EWSB are closely related.
In the context of a very simple extension of the Standard Model, the Inert Doublet Model, we show that dark matter could play a 
crucial role in the breaking of the electroweak symmetry. In this model, dark matter is the lightest component of an inert scalar doublet. The coupling of the latter with the Standard Model Higgs doublet breaks the electroweak symmetry at one-loop, {\it \`a la Coleman-Weinberg}. The abundance of dark matter, the breaking of the electroweak symmetry and the constraints from electroweak precision measurements can all be accommodated by imposing an (exact or approximate) custodial symmetry.  


\end{abstract}

 \maketitle

\section{Introduction}
One of the goals of the Large Hadron Collider is to elucidate the origin of Electroweak Symmetry Breaking (EWSB). In the framework of the Standard Model (SM), EWSB is expected to be due to the existence of a Brout-Englert-Higgs scalar doublet (Higgs doublet 
in the sequel)  which 
develops a non-zero vacuum expectation value ({\it vev}) at
 tree level. This necessitates  a  negative mass squared for the Higgs doublet, incidentally the only mass term allowed by the symmetries of the SM. 

An attractive possibility, proposed long ago by Coleman and Weinberg \cite{Coleman:1973jx,Sher:1988mj},  is that there is no tree level scalar mass altogether --perhaps because of some
 underlying conformal symmetry-- and that EWSB
 is
 caused by radiative corrections. However, appealing as it may be, this mechanism fails within the Standard Model. Because of the large negative contribution from 
top quark loop, either extra gauge bosons \cite{Coleman:1973jx,Hempfling:1996ht,Chang:2007ki}  or extra scalars \cite{Coleman:1973jx,Gildener:1976ih,Meissner:2006zh,Espinosa:2007qk,Foot:2007as} 
with large couplings must be added to the SM to get, within this approach,  a Higgs particle mass consistent with the experimental bound \footnote{Another possibility, that  we will not address here, 
is to consider the SM as an effective theory cutted off by a hard scale \cite{Hambye:1995fr}.}.

In an apparently different vein, the recent cosmological observations concur to indicate that dark matter exists and that it is even more abundant than ordinary matter~\cite{Spergel:2006hy,Seljak:2006bg}. The nature of the dark matter eludes us but a weakly interacting massive particle (WIMP) with mass around the electroweak scale, which was once in thermal equilibrium, would have a relic abundance
 consistent with observations. 

In this article we study a very simple
extension of the Standard Model that lies the origin of electroweak symmetry in the existence of a dark matter candidate and its $SU(2)$ partners and their one-loop contribution.
This 
scenario {\em \`a la Coleman Weinberg} can give a Higgs mass above the experimental value $M_H > 114.4$ GeV  together with a Dark Matter abundance consistent with cosmological observations, $\Omega h^2 \approx 0.12$ \cite{Yao:2006px}.

\section{The model}
The model we consider is a  two Higgs doublet extension of the SM, $H_1=(h^+ \, (h+iG_0)/\sqrt{2})^T$ and $H_2=(H^+ \, (H_0+iA_0)/\sqrt{2})^T$, 
together with a $Z_2$ symmetry such that 
all fields of the Standard Model and $H_1$  are even under $Z_2$ while
$
H_2\rightarrow - H_2. 
$
We assume that $Z_2$ is not spontaneously broken, {\em i.e.} that $H_2$ does 
not develop a {\it vev}. As there is no mixing between the doublets, $h$ plays the role of the usual Higgs particle. 
This very minimal extension of the SM  is called the Inert Doublet Model (IDM) because the extra (or inert)
 doublet does not couple to the quarks (and leptons in this version of the model).
 This feature is  consistent with the non-observation of Flavour Changing Neutral Currents.  

The IDM has been discussed long ago by Deshpande and Ma \cite{Deshpande:1977rw}. It contains a dark matter candidate in the form of either $H_0$ or $A_0$. 
This aspect has been considered in recent works, in particular in \cite{Cirelli:2005uq} as a minimal dark matter candidate, in \cite{Ma:2006km}  together with a mechanism to generate neutrino masses at one-loop and in \cite{Barbieri:2006dq}  as a framework with
 a heavy Higgs. This dark matter  have been further studied in \cite{Majumdar:2006nt,LopezHonorez:2006gr,Gustafsson:2007pc}.
\footnote{Variations on the IDM from various perspectives has been discussed in e.g. 
\cite{Gerard:2007kn,Pierce:2007ut,Lisanti:2007ec,Calmet:2006hs,Casas:2006bd,Kubo:2006yx,Hambye:2006zn,Ma:2006fn}.}

The most general renormalisable (CP conserving) potential of the model 
is 
\begin{eqnarray} 
\label{potential} 
V &=& \mu_1^2 \vert H_1\vert^2 + \mu_2^2 \vert H_2\vert^2  + \lambda_1 \vert H_1\vert^4 + 
 \lambda_2 \vert H_2\vert^4 \\ \nonumber 
&&\\
&&  + \lambda_3 \vert H_1\vert^2 \vert H_2 \vert^2 
 + \lambda_4 \vert H_1^\dagger H_2\vert^2 + {\lambda_5\over 2} \left[(H_1^\dagger H_2)^2 + h.c.\right] \nonumber
\end{eqnarray} 
with real quartic couplings.
The $SU(2) \times U(1)$ symmetry is broken by the vacuum expectation value of $H_1$, 
$\langle H_1\rangle = {v/\sqrt 2}
$ 
with $v = -\mu_1^2/\lambda_1 = 246$ GeV 
while, assuming $\mu_2^2 > 0$, 
$
\langle H_2\rangle = 0. 
$ 
The mass of the 
 Higgs boson, $h$,
 is 
\begin{equation}
m_h^2 = \mu_1^2+ 3 \lambda_1  v^2 \equiv - 2 \mu_1^2 = 2 \lambda_1 v^2 
\label{hmass}
\end{equation}
while the mass of the 
charged, $H^+$, and two neutral, $H_0$ and $A_0$, components of the field $H_2$ are given by 
\begin{eqnarray} 
m_{H^+}^2 &=& \mu_2^2 + \lambda_3 v^2/2\cr 
m_{H_0}^2 &=& \mu_2^2 +  (\lambda_3 + \lambda_4 + \lambda_5) v^2/2\cr 
m_{A_0}^2 &=& \mu_2^2 +  (\lambda_3 + \lambda_4 - \lambda_5) v^2/2. 
\label{masses} 
\end{eqnarray} 
Various limits are of interest.
There is a Peccei-Quinn symmetry if  $\lambda_5 =0$,  with $m_{H_0}=m_{A_0}$. This limit is however disfavoured 
by constraints from dark matter direct detection experiments \cite{Barbieri:2006dq,Cirelli:2005uq,LopezHonorez:2006gr}.
In the limit $\lambda_4= \lambda_5$, or in the twisted case $\lambda_4=-\lambda_5$ \cite{Gerard:2007kn}, there is a custodial $S0(3)$ symmetry, with $m_{H^\pm} = m_{A_0}$ or $m_{H^\pm} = m_{H_0}$, respectively. 
We will come back to the custodial symmetry when we will discuss constraints from LEP precision measurements.  Following \cite{Barbieri:2006dq} we parameterise the contribution from 
symmetry breaking to the mass of $H_0$ and $A_0$ by $\lambda_{L,S}=\lambda_3 + \lambda_4 \pm 
\lambda_5$ (which are also the coupling constants between the Higgs field $h$
  and our dark matter candidates $H_0$ or $A_0$ respectively).

For appropriate quartic couplings, either $H_0$ or $A_0$ is the 
lightest component of the $H_2$ doublet and, in absence of other lighter $Z_2$-odd fields, 
either one is a candidate for dark matter. 
There are {\em a priori} two distinct dark matter mass scales which have a relic density  consistent with WMAP data: 
a low-mass one, $M_{DM} \lesssim 75$ GeV, below the threshold for $W$ pair production, and a large mass one,  $M_{DM} \gsim 400$ GeV \cite{Barbieri:2006dq,Cirelli:2005uq,LopezHonorez:2006gr}. 
The former case is the most promising one from
the point of view  of direct and/or indirect detection \cite{LopezHonorez:2006gr,Gustafsson:2007pc}. 
In this case the DM relic abundance is dictated by a) annihilation of DM into the Higgs, whose efficiency depends on $M_h$ and $\lambda_L$ or $\lambda_S$, b)
 annihilation into a $W^\pm$ pair, as $M_{DM}$ gets closer to  $M_{W^\pm}$ and c) coannihilation of $H_0$ and  $A_0$ (resp.~of DM and $H^\pm$) into a $Z$ boson (resp.~$W^\pm$), 
if the mass splitting between $H_0$ and $A_0$ (resp.~between DM and $H^\pm$) is, roughly speaking, close to the freeze-out temperature  $T_{fo}\sim M_{DM}/20$.  

\section{One-loop radiative corrections}


We now consider one-loop corrections to the Higgs effective potential,
which  is  given by the usual expression
\begin{equation}
V_{\hbox{eff}}(h)=\mu_1^2 \frac{h^2}{2} +\lambda_1 \frac{h^4}{4} + \frac{1}{64 \pi^2}
 \sum_in_i   m_i^{4}\big(ln\frac{m^{2}_{i}}{\mu^2}-c_i\big)
\end{equation}
where $n_i=\{1,1,1,1,2,2,-12,2,4\}$ is the number of degrees of freedom of each species $i=\{h,H_0,G_0,A_0,h^\pm,H^\pm,t,Z,W^\pm\}$ 
which couples to the Higgs boson with tree level mass (\ref{hmass}) and (\ref{masses}) while  $m^{2}_{G_0}= m^{2}_{h^\pm}= \mu_1^2+\lambda_1 v^2$, $m_t^2=g_t^2 v^2/2$, $m_W^2=g^2 v^2/2$ and $m_Z^2=(g^2+g'^2)v^2/2$. The constant is $c_i=3/2$ for all scalars and fermions and $c_i=5/6$ for all gauge bosons. The gauge bosons loops are given here for completeness. However, as their effects are generically small, we will neglect their contribution in the sequel. 

Imposing that the effective potential has an extremum for $\langle h\rangle=v=246$ GeV, the Higgs mass at one-loop is given by
\begin{eqnarray}
M_h^2&=&\frac{d^2 V_{\hbox{eff}}}{d h^2}= m_h^2 +\frac{1}{32 \pi^2} \, \Big[ 6 \lambda_1 f(m_h^2) +\lambda_L f(m^2_{H_0})\nonumber \\
&+&2 \lambda_1 f(m^2_{G_0})
+\
\lambda_S f(m^2_{A_0})+4 \lambda_1 f(m^2_{h^+})+2 \lambda_3 f(m^2_{H^+})\nonumber \\
&+&36 \lambda_1^2 h^2 \log \frac{m_h^2}{\mu^2}
+\lambda_L^2 h^2 \log \frac{m_{H_0}^2}{\mu^2}
+4 \lambda_1^2 h^2 \log \frac{m^2_{G_0}}{\mu^2}\nonumber \\
&+&\lambda_S^2 h^2 \log \frac{m_{A_0}^2}{\mu^2}
+8 \lambda_1^2 h^2 \log \frac{m^2_{h^+}}{\mu^2}
+2 \lambda_3^2 h^2 \log \frac{m^2_{H^+}}{\mu^2} 
\nonumber \\
&-&36 g_t^2 h^2 f(m_t^2)-12 g_t^4 h^2 \Big]\Big|_{\langle h\rangle=v}
\label{mh}
\end{eqnarray}
with $f(m^2)=m^2 (\log (m^2/\mu^2)-1)$.

Since $H_2$ has no vacuum expectation value, there is no mixing between the scalars and it is straightforward to
 compute the contribution of one-loop corrections to the mass of the other scalars from the second derivative of the effective potential around the Higgs {\em vev} (see for instance \cite{Peskin:1995ev}, section 11.6). 
This still requires to keep track of  the dependence of the propagators on $h$, $H_0$, $A_0$ and $H^\pm$ though.
The fact that there is no mixing also means that the extremum is necessarily a minimum if all masses
are positive.
The result is, using the $\overline{MS}$ prescription,
\begin{eqnarray}
M^2_{H_0}&\equiv& \frac{\partial^2 V_{\hbox{eff}}}{\partial H_0^2}=  m^2_{H_0} + \frac{1}{32 \pi^2} \Big[ \lambda_L f(m_h^2) +6 \lambda_2 f(m^2_{H_0})  \nonumber \\
+&&\hspace{-6mm}
\lambda_S f(m^2_{G_0})+2 \lambda_2 f(m^2_{A_0})+2 \lambda_3 f(m^2_{h^+})+4 \lambda_2 f(m^2_{H^+})\nonumber \\
-&&\hspace{-6mm} 2\lambda_L^2 v^2 g(m_h^2,m^2_{H_0}) -2 \lambda^2_5 v^2 g(m^2_{G_0},m^2_{A_0}) \nonumber \\
-&&\hspace{-6mm} (\lambda_4+\lambda_5)^2 v^2 g(m^2_{h^+},m^2_{H^+})
\Big] \Big|_{\langle h\rangle=v}
\label{meta0} 
\\
M^2_{A_0}&\equiv& \frac{\partial ^2 V_{\hbox{eff}}}{\partial  A_0^2}= m_{A_0}^2 +  \frac{1}{32 \pi^2} \Big[ \lambda_S f(m_h^2) +2 \lambda_2 f(m^2_{H_0})\nonumber \\
+&&\hspace{-6mm}
\lambda_L f(m^2_{G_0})+6 \lambda_2 f(m^2_{A_0})+2 \lambda_3 f(m^2_{h^+})+4 \lambda_2 f(m^2_{H^+})\nonumber \\
-&&\hspace{-6mm} 2 \lambda_S^2 v^2 g(m_h^2,m^2_{A_0}) -2 \lambda^2_5 v^2g(m^2_{G_0},m^2_{H_0}) \nonumber \\
-&&\hspace{-6mm} (\lambda_4-\lambda_5)^2 v^2 g(m^2_{h^+},m^2_{H^+}) \Big] \Big|_{\langle h\rangle=v}
\label{mA0}
\\
M^2_{H^\pm}&\equiv& \frac{\partial^2 V_{\hbox{eff}}}{\partial H^+ \partial H^-} = m^2_{H^\pm} +  \frac{1}{32 \pi^2} \Big[ \lambda_3 f(m_h^2) +2 \lambda_2 f(m^2_{H_0})\nonumber \\
+&&\hspace{-6mm}
\lambda_3 f(m^2_{G_0})+2 \lambda_2 f(m^2_{A_0})+2(\lambda_3+\lambda_4)  f(m^2_{h^+})\nonumber \\
+&&\hspace{-6mm} 8 \lambda_2 f(m^2_{H^+}) - {1\over 2} (\lambda_4+\lambda_5)^2 v^2 g(m^2_{h^+},m^2_{H_0}) 
\nonumber \\
-&&\hspace{-6mm} 2\lambda_3^2 v^2 g(m_h^2,m^2_{H^+})  -\frac{1}{2} (\lambda_4-\lambda_5)^2 v^2 g(m^2_{h^+},m^2_{A_0})
\Big] \Big|_{\langle h\rangle=v}.
\label{mH}
\end{eqnarray}
with $g(m_1^2,m_2^2)=[f(m_1^2)-f(m_2^2)]/(m_2^2-m_1^2)$.

In all these expressions, we take $\mu = m_t=172.5$ GeV. In principle, a change in the renormalisation scale is compensated by the scale dependence of the running quartic couplings. 
However implementing this is a lengthy task since their beta functions mix the different couplings (cf Eq.(61) of appendix B in \cite{Barbieri:2006dq}). At 
the present exploratory stage we simply neglect the running of the couplings.

\section{EWSB and Dark Matter}

We first focus on the physically appealing case of vanishing mass terms, or conformal limit $\mu_1=\mu_2=0$. For the sake of completeness, we will comment on the  case $\mu_2,\mu_1\neq 0$ at the end of this section. In the conformal limit there are  three important constraints:

1) \underline{EWSB}. 
The general strategy is simple. The contribution of at least some of the loops with $H_2$ particles must be large enough to compensate the large, negative, contribution 
of the top quark. This requires that at least one of  the $\lambda_{3-5}$ couplings 
must be large and positive. This will inevitably drive some of the scalar particle masses in the few hundred GeV range. Imagine  that  EWSB is driven by loop corrections of 
$H^\pm$ and $A_0$, with $\lambda_3\simeq \lambda_S$.  Since these particles represent together only 3 degrees of freedom whereas there are $12$ for the top quark, the $\lambda_{3,S}$ contribution is relevant only provided 
 $\lambda_{3,S} \gsim 2 g_t^2$. Asking that their contribution
 is large enough for the Higgs mass to be above $\sim 115$ GeV requires  $\lambda_{3,S} \gsim 5 g^2_t$, approximately. 
This gives $M_{H^\pm,A_0} \gsim 380$ GeV.  

2) \underline{Low DM mass}. 
In general (see {\em e.g.}~\cite{LopezHonorez:2006gr}) the mass of DM comes  from both $\mu_2$ and  the coupling to the Higgs. If $M_{DM} > M_{W}$, 
the dominant process for the relic abundance of $H_0$ is the annihilation into  $W^\pm$ and $Z$ pairs. If
$\lambda_L=0$ (or $\lambda_L=0$), the
cross-section scales like $1/M_{DM}^2$ (this is expected on general grounds \cite{Griest:1989wd}) and, for a sufficiently large mass, $M_{DM} \gsim 400$ GeV, 
the abundance is consistent with observations.  
However, if we increase the coupling to the Higgs, it turns out that the DM annihilation cross-section increases  and so the DM abundance decreases. 
 (This behaviour is precisely analogous to that  
of a heavy SM neutrino, whose annihilation cross-section also increases for large neutrino masses \cite{Enqvist:1988we,Kainulainen:2002pu}.) 
Consequently, if $\mu_2=0$ and all the mass comes from the coupling to the Higgs,
 the abundance of a heavy DM
 is  much smaller than observation and the only viable possibility for dark matter  is if $M_{DM} < M_{W}$. 

3) \underline{Electroweak precision measurement}. Since at least one of the components of the inert doublet must be very heavy to break the electroweak symmetry while the DM candidate 
 must be lighter than $M_{W}$, we have to face the constraints from Electroweak Precision Measurements \footnote{In \cite{Barbieri:2006dq} the mass splitting 
must be kept small $\lsim 15$ GeV because the Higgs is assumed to be very heavy $M_h \gsim 500$ GeV and the abundance is dictated by coannihilation. This does not apply in our case because
 annihilation goes through the Higgs.}. 
A doublet with large mass splitting will contribute to the SM $\rho$ parameter or, equivalently, to the Peskin-Takeuchi  $T$ parameter. 
At one-loop 
\begin{equation}
\label{T}
\Delta T  = {1\over 32 \pi^2 \alpha v^2} \left[ f(M_{H^\pm},M_{H_0}) + f(M_{H^\pm},M_{A_0}) - f(M_{A_0}, M_{H_0})\right]
\end{equation}
with $f(m_1,m_2) = (m_1^2 + m_2^2)/2 - m_1^2 m_2^2/(m_1^2 - m_2^2) \ln(m_1^2/m_2^2)$ \cite{Barbieri:2006dq}. To give an idea, the contribution from $M_{H^\pm} \sim 450$ GeV and $M_{DM} \sim 75$ GeV tree level masses gives
$\Delta T \sim 1$ while
electroweak precision measurements impose $\vert \Delta T\vert \lsim 0.2$. Since the inert doublet is massless at tree level, strictly speaking $\Delta T$ vanishes at one-loop. 
 Nevertheless we should take the issue seriously as the large mass differences we are after will inevitably 
 give a large contribution to the gauge boson mass splitting, be it beyond one-loop order. 
There is however a nice and  painless cure to this problem: as a quick 
inspection of Eq. (\ref{T}) reveals, if
 either $H_0$ or $A_0$ is degenerate with $H^\pm$, the contribution of the inert doublet to the $\Delta T$ parameter vanishes identically. Physically, 
this is due to the existence of a custodial symmetry in the limit $M_{H^\pm}=M_{A_0}$ or $M_{H^\pm}=M_{H_0}$ (i.e. $\lambda_4 =\pm \lambda_5$). \footnote{
Notice that the hypothesis of custodial symmetry together with the $Z_2$ symmetry  automatically gives CP conservation in the scalar 
sector \cite{Gerard:2007kn}.
CP violation ({\em i.e.} $\lambda_5$ complex) is potentially dangerous for dark matter. It leads to $H_0$-$A_0$ mixing and thus to the possibility of spin-independent direct detection through $Z$-exchange. Unless the phase of $\lambda_5$ is tiny, this induces a far too large direct detection rate for WMAP abundances.}.
Technically, an exact or approximate custodial symmetry does not only 
avoid large corrections to the $T$ parameter. It also 
implies that it is no fine tuning to take, for instance, the $DM$ particle  to be lighter than 
the other  components of the inert doublet (i.e. $\lambda_L$ or $\lambda_S$ much different from the other quartic couplings) as required by the EWSB and DM constraints. We think
 that this feature holds beyond one-loop order. 

From the three constraints above, we can now consider four cases (see Table 1). Case I corresponds to a light $H_0$ and to two heavy, nearly degenerate $A_0$ and $H^\pm$ ({\em i.e.} $m_{H_0} << m_{A_0}\simeq m_{H^+}$ or $\lambda_L < < \lambda_S \simeq \lambda_3$). 
Case II has a reversed hierarchy, {\em i.e.}   
$m_{H_0} \lesssim m_{H^+}   << m_{A_0} $ or $  \lambda_L   \lesssim \lambda_3 << \lambda_S $). The two last corresponds to 
 $A_0$ as the DM candidate, with $m_{A_0} << m_{H_0}\simeq m_{H^+}$ (case III) and  
$m_{A_0} \lesssim  m_{H^+} << m_{H_0}$ (case IV).
Cases III and IV can be obtained from cases I and II simply by switching
$H_0$ with $A_0$, i.e.~$\lambda_5$ with $-\lambda_5$. This leaves the relic density  unchanged, so that Table 1 is relevant for these cases too.
 
All the examples of Table 1 have a DM abundance in agreement with WMAP data \footnote{The relic abundance was computed using Micromegas2.0 \cite{Belanger:2006is}}. 
As announced, we observe  that some of the quartic couplings must be large. Also, in all the working cases the DM mass 
is below $M_W$. In Case I (similarly case III), the DM abundance is determined by its annihilation through the Higgs particle only and thus depends on $M_h$ and the 
effective trilinear $hH_0H_0$ coupling, {\it i.e.}
$\lambda_L^{eff}=\frac{1}{v}\partial^3 V_{eff}/ \partial h \partial^2 H_0\equiv \frac{1}{v}\partial M^2_{H_0}/ \partial v$ at one-loop. 
For various, albeit large, couplings we found  the correct abundance for DM masses in the range  $M_{H_0}\sim (10-72)$ GeV.  Below this range, the Higgs mediated annihilation is too suppressed.
We remark that the values of $M_{H_0}$ consistent
with DM and EWSB can be below the ones found in the tree level analysis of \cite{LopezHonorez:2006gr}. This is because the one-loop contributions to $\lambda_L^{eff}$ 
can  be sizeable, {\em i.e.} for the same mass, the DM particle can be more strongly coupled to the Higgs than it is at tree level. 
In case II (resp.~case IV) coannihilation through the $W^+$ can play a role if the $H^+-H_0$ (resp.~$H^+-A_0$) splitting is not too large. Notice that the masses of $H^\pm$ quoted in Table 1 are consistent with collider data  because the $H^+$ does not couple to fermions, is short lived and, if $M_{H^\pm} > M_Z/2$,  does not contribute to the width of the Z boson.
Notice also that, unlike in cases I and III where it plays little role for DM, in cases II and IV the custodial symmetry  can only
be  approximate, otherwise the coannihilation (and direct detection) cross-sections would be  too large to be consistent with observations. 
Notice finally that Cases II and IV require larger quartic coupling because they involve only one heavy degree of freedom in EWSB instead of three in Cases I and III. 

Imposing the quartic couplings $\lambda_{3,L,S}$ to be smaller than {\em e.g.} $2 \pi$ or $4 \pi$ gives $M_h \lsim 80$ GeV or $M_h\lsim 175$ GeV  in Cases II and IV while for Cases I and III we have $M_h \lsim 150$ GeV or $M_h \lsim 350$ GeV. We have checked that these $M_h$ bounds can be saturated, keeping $\Omega_{DM} \sim 0.12$.
All these numbers are clearly tentative as the quartic couplings are quite large and, even if we are still in the perturbative regime, see e.g. Eqs.~(16-18) of \cite{Barbieri:2006dq}, higher order corrections could be important. However we do not think they would dramatically change the picture  drawn here.


\begin{table}
\label{tab1}
\begin{tabular}{|c |c|c|c|c|c||c|c|c|c||c|c|c|}\hline
 & $\lambda_1$ &$\lambda_2$  & $\lambda_3$& $\lambda_4$ &$\lambda_5$ & $M_h$ & $M_{H_0}$ &  $M_{A_0}$ & $M_{H^\pm}$  & $h_{BR}$ & $W_{BR}$\\
\hline\hline
I & -0.11 & 0 & 5.4 & -2.8 & -2.8 & 120  & 12 & 405 & 405 & 100\% & 0\%\\
\hline
I & -0.11&-2 & 5.4 & -2.7 & -2.7 & 120 & 43 & 395 & 395 & 100\% & 0\%\\
\hline
I & -0.11 &-3&  5.4 &  -2.6 & -2.6 & 120 & 72 & 390 & 390 & 94\% & 6 \% \\
\hline
I & -0.30 & 0 & 7.6 & -4.1 & -4.1 & 180 & 12 & 495 & 495 & 100\% & 0 \% \\
\hline
I & -0.30 & -2.5 & 7.6 & -3.8 & -3.8 & 180 & 64 & 470 & 470 & 100\% & 0 \% \\
\hline
II & -0.18 & -3&  -0.003 & 4.6 & -4.7 & 120 & 39 & 500 & 55 & 100\% & 0 \% \\
\hline
II & -0.29 & -5&  -0.07 & 5.5 & -5.53 & 150 & 54 & 535 & 63 & 0\% & 100 \% \\
\hline
\end{tabular}
\caption{Instances of parameters with WMAP DM abundance.  
Also given are the relative contribution of Higgs mediated annihilation ($h_{BR}$) and gauge  processes ($W_{BR}$).}   
\end{table}
 

It should be clear from these results that, even if $\mu_1,\,\mu_2 \neq 0$, the existence of DM around the electroweak scale 
could have startling effects 
on EWSB. With respect to the conformal limit, larger (smaller) quartic coupling than in the conformal case should be considered for $\mu_2^2>0$ (resp. $\mu_2^2 <0$).

Instead of an inert doublet we could consider higher-dimensional inert Higgs multiplets. The case of a scalar singlet has already been considered to induce 
EWSB \cite{Espinosa:2007qk,Patt:2006fw}. In our opinion, in the latter case the connection discussed in the present paper  would be looser since it could not be the 
same object that drive both the EWSB and has the right relic DM abundance.  Assuming their masses to be around the EW scale, several singlets with large (for EWSB) and small couplings (for DM \cite{McDonald:2001vt}) would presumably be 
necessary. 

\section{Summary}

We have shown that Dark Matter in the form of the lightest neutral component of a single inert doublet could be responsible for EWSB. 
We have met essentially three constraints.  A large quartic coupling is necessary to drive EWSB. One quartic coupling must be small to have a  DM particle mass below $M_W$. 
Finally a small mass splitting of either the $A_0$ or $H_0$ with $H^\pm$ is required to confront Electroweak Precision Measurements.
All these conditions can be satisfied naturally if an exact or approximate custodial symmetry is assumed. 
As a result of all constraints we get the bound on the mass of the Higgs $M_h \lesssim 350$~GeV while the mass of dark matter is in the range $M_{DM} \sim (10-72)$ GeV. 
Such a  DM candidate is in a range of couplings that  makes it accessible to both direct (ZEPLIN) and indirect (GLAST) future searches (cf Figure 5 of \cite{LopezHonorez:2006gr}). 

Since the quartic couplings are quite large, the results of the present paper are probably only tentative. Nevertheless we do expect that the breaking of the Electroweak Symmetry with a  WIMP Dark Matter candidate is a feature of the Inert Doublet Model which will survive further investigations.

\section*{Acknowledgement}
 We thank J.-M. G\'erard for useful discussions. This work is supported by the FNRS, the IISN and the Belgian Federal Science Policy (IAP VI/11). Preprint ULB-TH/07-26.


\begin{thebibliography}{99}

\bibitem{Coleman:1973jx}
  S.~R.~Coleman and E.~Weinberg,
  Phys.\ Rev.\  D {\bf 7} (1973) 1888.


 
  \bibitem{Sher:1988mj} 
  M.~Sher, 
  Phys.\ Rept.\  {\bf 179}, 273 (1989). 

\bibitem{Hempfling:1996ht}
  R.~Hempfling,
  Phys.\ Lett.\  B {\bf 379} (1996) 153.

\bibitem{Chang:2007ki}
  W.~F.~S.~Chang, J.~N.~Ng and J.~M.~S.~Wu,
  arXiv:hep-ph/0701254.

\bibitem{Gildener:1976ih}
  E.~Gildener and S.~Weinberg,
  Phys.\ Rev.\  D {\bf 13}, 3333 (1976).

\bibitem{Meissner:2006zh}
  K.~A.~Meissner and H.~Nicolai,
  Phys.\ Lett.\  B {\bf 648}, 312 (2007).

\bibitem{Espinosa:2007qk}
  J.~R.~Espinosa and M.~Quiros,
  arXiv:hep-ph/0701145.

\bibitem{Foot:2007as}
  R.~Foot, A.~Kobakhidze and R.~R.~Volkas,
  arXiv:0704.1165 [hep-ph].

\bibitem{Hambye:1995fr}
  T.~Hambye,
  Phys.\ Lett.\  B {\bf 371}, 87 (1996).

  

\bibitem{Spergel:2006hy}
  D.~N.~Spergel {\it et al.},
  arXiv:astro-ph/0603449.
\bibitem{Seljak:2006bg}
  U.~Seljak, A.~Slosar and P.~McDonald,
  JCAP {\bf 0610} (2006) 014.

\bibitem{Yao:2006px}
  W.~M.~Yao {\it et al.}  [Particle Data Group],
  J.\ Phys.\ G {\bf 33} (2006) 1.

\bibitem{Deshpande:1977rw}
  N.~G.~Deshpande and E.~Ma,
  Phys.\ Rev.\ D {\bf 18} (1978) 2574.



 


\bibitem{Cirelli:2005uq} 
  M.~Cirelli, N.~Fornengo and A.~Strumia, Nucl. Phys. {\bf B753} (2006) 178.



\bibitem{Ma:2006km} 
  E.~Ma, 
  Phys.\ Rev.\ D {\bf 73} (2006) 077301. 


 
\bibitem{Barbieri:2006dq} 
  R.~Barbieri, L.~J.~Hall and V.~S.~Rychkov, 
  Phys.\ Rev.\ D {\bf 74} (2006) 015007 
  [arXiv:hep-ph/0603188]. 




\bibitem{Majumdar:2006nt} 
  D.~Majumdar and A.~Ghosal, 
  arXiv:hep-ph/0607067. 
 

\bibitem{LopezHonorez:2006gr}
  L.~Lopez Honorez, E.~Nezri, J.~F.~Oliver and M.~H.~G.~Tytgat,
  JCAP {\bf 0702} (2007) 028

\bibitem{Gustafsson:2007pc}
  M.~Gustafsson, E.~Lundstrom, L.~Bergstrom and J.~Edsjo,
  arXiv:astro-ph/0703512.

\bibitem{Gerard:2007kn}
  J.~M.~G\'erard and M.~Herquet,
  arXiv:hep-ph/0703051.

\bibitem{Pierce:2007ut}
  A.~Pierce and J.~Thaler,
  arXiv:hep-ph/0703056.

\bibitem{Lisanti:2007ec}
  M.~Lisanti and J.~G.~Wacker,
  arXiv:0704.2816 [hep-ph].

\bibitem{Calmet:2006hs}
  X.~Calmet and J.~F.~Oliver,
  Europhys.\ Lett.\  {\bf 77} (2007) 51002.


 
\bibitem{Casas:2006bd} 
  J.~A.~Casas, J.~R.~Espinosa and I.~Hidalgo, 
  arXiv:hep-ph/0607279. 
 


 
\bibitem{Kubo:2006yx} 
  J.~Kubo, E.~Ma and D.~Suematsu,  Phys. Lett. {\bf B642} (2006) 18.
 
\bibitem{Hambye:2006zn} 
  T.~Hambye, K.~Kannike, E.~Ma and M.~Raidal, Phys. Rev. {\bf D75} (2007) 095003.
 
\bibitem{Ma:2006fn} 
  E.~Ma, 
  Mod.\ Phys.\ Lett.\ A {\bf 21} (2006) 1777.
 
\bibitem{Peskin:1995ev}
  M.~E.~Peskin and D.~V.~Schroeder,
{\it  Reading, USA: Addison-Wesley (1995) 842 p}



 

\bibitem{Jungman:1995df}
  G.~Jungman, M.~Kamionkowski and K.~Griest,
  Phys.\ Rept.\  {\bf 267} (1996) 195.




 
\bibitem{Griest:1989wd}
  K.~Griest and M.~Kamionkowski,
  Phys.\ Rev.\ Lett.\  {\bf 64} (1990) 615.




\bibitem{Enqvist:1988we}
  K.~Enqvist, K.~Kainulainen and J.~Maalampi,
  Nucl.\ Phys.\ B {\bf 317} (1989) 647.

\bibitem{Kainulainen:2002pu}
  K.~Kainulainen and K.~A.~Olive,
  arXiv:hep-ph/0206163.

\bibitem{Belanger:2006is}
  G.~Belanger, F.~Boudjema, A.~Pukhov and A.~Semenov,
  Comput.\ Phys.\ Commun.\  {\bf 176} (2007) 367.

\bibitem{Patt:2006fw}
  B.~Patt and F.~Wilczek,
  arXiv:hep-ph/0605188.

\bibitem{McDonald:2001vt}
  J.~McDonald,
  Phys.\ Rev.\ Lett.\  {\bf 88} (2002) 091304.

  






\end{thebibliography}

\end{document}